\documentclass[12pt,preprint]{aastex}
\newcommand{\kms}{km~s$^{-1}$}
\usepackage{natbib}
\usepackage{graphicx}
\usepackage{amssymb}
\begin{document}

\title{The HDO/H$_2$O ratio in gas in the inner regions of a low-mass protostar}
\author{Jes K. J{\o}rgensen}
\affil{Centre for Star and Planet Formation, Natural History Museum of
Denmark, University of Copenhagen, {\O}ster Voldgade 5-7, DK-1350
Copenhagen K., Denmark}
\email{jes@snm.ku.dk}
\and
\author{Ewine F. van Dishoeck}
\affil{Leiden Observatory, Leiden University, PO Box 9513, NL-2300 RA
Leiden, The Netherlands \\ \& Max-Planck Institut f\"ur extraterrestrische
Physik, Giessenbachstrasse, D-85748 Garching, Germany}
\email{ewine@strw.leidenuniv.nl}

\begin{abstract}
  The HDO/H$_2$O abundance ratio is thought to be a key diagnostic on
  the evolution of water during the star- and planet-formation process
  and thus on its origin on Earth. We here present
  millimeter-wavelength high angular resolution observations of the
  deeply embedded protostar NGC~1333-IRAS4B from the Submillimeter
  Array targeting the $3_{12}-2_{21}$ transition of HDO at 225.6~GHz
  ($E_u$~=~170~K). We do not (or only very tentatively) detect the HDO
  line toward the central protostar, contrasting the previous
  prominent detection of a line from another water isotopologue,
  H$_2^{18}$O, with similar excitation properties using the IRAM
  Plateau de Bure Interferometer. The non-detection of the HDO line
  provides a direct, model independent, upper limit to the HDO/H$_2$O
  abundance ratio of 6$\times 10^{-4}$ (3$\sigma$) in the warm gas
  associated with the central protostar. This upper limit suggests
  that the HDO/H$_2$O abundance ratio is not significantly enhanced in
  the inner $\approx$~50~AU around the protostar relative to what is
  seen in comets and Earth's oceans and does not support previous
  suggestions of a generally enhanced HDO/H$_2$O ratio in these
  systems.
\end{abstract}


\maketitle

\section{Introduction}\label{introduction}
One of the key questions concerning the formation of planets, and our
own Earth in particular, is how water has evolved on its way from
collapsing molecular cloud cores, to protoplanetary disks and
\emph{eventually} Earth's oceans. In cold and quiescent regions, the
gas-phase water abundance is low, only $10^{-9}-10^{-8}$ or less
\citep[e.g.,][]{bergin02h2o,caselli10wish}, but in regions forming
low-mass protostars with intense heating or active shocks, its
abundance can reach 10$^{-4}$ with respect to H$_2$ - comparable to or
higher than that of CO \citep[e.g.,][]{kristensen10wish}. Thus, in the
earliest prestellar stages and in large parts of the dense envelopes
around low-mass protostars, H$_2$O is largely frozen out and the
dominant constituent of the icy mantles of dust grains. It is thought
that the amount of deuterated relative to non-deuterated water is
established in these grain mantles. Observations show that this
abundance ratio is enhanced in comets ($\sim 3-4\times 10^{-4}$; e.g.,
\citealt{bockeleemorvan98,villanueva09}) and Earth's oceans ($\sim
1.5\times 10^{-4}$; e.g., \citealt{robert00} and references therein)
relative to primordial D/H ratio in the protosolar nebula
($\sim$~1.5$\times 10^{-5}$; \citealt{geiss98,linsky03}). Determining
when and where this ratio is established is thus a potentially
important way to trace the evolution of water in the different stages
of star formation. In particular, it might help determine what
fraction of water was brought to Earth at later times through cometary
impacts compared to being accreted during the formation of the planet
itself \citep[e.g.,][]{morbidelli00,raymond04} and whether water
undergoes signficant processing in warm regions of protostellar
envelopes or disks before reaching newly formed planets.

Previous attempts at measuring the ratio of HDO and H$_2$O in low-mass
protostars have had mixed success. Direct attempts to detect HDO in
the solid phase at infrared wavelengths have provided upper limits of
[HDO]/[H$_2$O] varying from 0.005 to 0.02 in a sample of protostars
\citep{parise03}, which are not strong constraints compared to the two
orders of magnitude lower abundance ratios observed in, e.g., the
cometary systems about two orders of magnitude lower. Measurements at
(sub)millimeter wavelengths could potentially be sensitive to lower
abundances of HDO in the gas-phase in regions where the ices have
evaporated off the dust grains and thus trace their compositions
indirectly, but such measurements are complicated due to the
achievable angular resolution using single-dish radio telescopes and
have led to discrepant results. As an example, \cite{stark04} and
\cite{parise05} analyzed the HDO emission toward the deeply embedded
protostar, IRAS~16293-2422 compared to previous ISO and SWAS
observations of non-deuterated H$_2$O. Both presented detailed models
for the HDO abundance structure with significantly different results:
\citeauthor{parise05} found evidence for a drop in the [HDO]/[H$_2$O]
ratio from 0.03 in the hot gas in the inner envelope (i.e.,
significantly enhanced above the cometary values) to $<$0.002 in the
cold gas on larger scales, whereas \citeauthor{stark04} in contrast
found that the abundance of HDO was best described as being constant
throughout the envelope and with a [HDO]/[H$_2$O] ratio of 2$\times
10^{-4}$, comparable to the cometary value, in the part of the
envelope with $T > 14$~K. Part of this discrepancy may arise due to
the complexities in comparisons between ground-based single-dish
observations of HDO and space-based observations of H$_2$O probing
signicantly different spatial scales - and lines excited in different
regions of the protostellar environment.

Our detection of thermal H$_2^{18}$O emission toward the deeply
embedded Class 0 low-mass protostar, NGC~1333-IRAS4B ($d=250$~pc),
with the IRAM Plateau de Bure Interferometer \citep{iras4b_h2o} makes
it possible to revisit the discussion of the origin of water on Earth
- namely where the relative abundances of deuterated and
non-deuterated are established.  In this letter we present a search
for the 225.9~GHz $3_{12}-2_{21}$ transition of the HDO isotopologue
at high-angular resolution (2.5--3$''$; corresponding to 625--700~AU)
using the Submillimeter Array (SMA). This transition is close in
energy to the H$_2^{18}$O $3_{13}-2_{20}$ transition from
\cite{iras4b_h2o} ($E_{\rm up} = 170$~K vs. 204~K) and a
straightforward comparison can thus be made between these species with
minimal uncertainties due to their excitation properties.

\section{Observations}
We observed NGC~1333-IRAS4B (IRAS4B in the following;
$\alpha$=03\fh29\fm12.00\fs, $\delta$=+31\fd13\arcmin08\farcs1 [J2000]
\citealt{prosacpaper}) using the Submillimeter Array (SMA;
\citealt{ho04}) on 2008 October 31. The receivers were tuned to the
HDO $3_{1,2}-2_{2,1}$ transition at 225.89672~GHz and the SMA
correlator assigned a chunk with 512 channels centered on this line
corresponding to a 0.27~\kms\ spectral resolution over about 100~MHz
of the chunk. The chunk was placed so that the upper sideband covered
the D$_2$CO transition at 236.102128~GHz with similar spectral
resolution. For the remaining SMA passband 128 channels were allocated
in each chunk giving a uniform spectral coverage of 1.1~\kms\ from
225.0 GHz to 227.0~GHz (LSB) and 235.0 to 237.0~GHz (USB). The source
was observed in the compact configuration of the SMA, which at the
time of observations had 7 antennae operational providing baselines
ranging from 7 to 60~k$\lambda$.

The data were calibrated using the SMA version of the MIR package and
imaged using Miriad. The calibration followed the standard approach:
the absolute flux calibration was established through observations of
Uranus, the bandpass by observations of the strong quasar 3c454.3 and
the complex gains by regular observations of the nearby strong quasars
3c84 and 3c111 (approximately 5.2 and 4.7~Jy at 1.33~mm,
respectively). Integrations with clearly deviating amplitudes and/or
phases were flagged and the continuum was subtracted prior to Fourier
transformation of the line data. With natural weighting the resulting
beam size is 3.1\arcsec $\times$ 2.5\arcsec at a position angle of
-66$^\circ$; the field of view is 52\arcsec\ (HPBW) at 1.33~mm. The
resulting RMS noise level is 58~mJy~beam$^{-1}$~channel$^{-1}$ for the
line data with the 0.27~\kms\ spectral resolution using natural
weighting.

\section{Results}
Both IRAS4B and its nearby companion IRAS4B$'$ are clearly detected in
the continuum data. Table~\ref{cont_table} lists the results of
elliptical Gaussian fits to the two sources: both are resolved with
fluxes in agreement with the results from \cite{prosacpaper} within
the absolute flux calibration uncertainty of 20\%.
\begin{table}\centering
  \caption{Parameters for IRAS4B and IRAS4B$'$ from elliptical Gaussian fits to their continuum emission.}\label{cont_table}
\begin{tabular}{lll}\hline\hline
                              & IRAS4B                   & IRAS4B$'$ \\ \hline
Flux                          &                1.021~Jy     &      0.364~Jy     \\
R.A. (J2000)                  &          03:29:12.0    & 03:29:12.8   \\
DEC (J2000)                   &          31:13:08.1    & 31:13:06.9  \\
Extent$^{a}$     &      1.7\arcsec$\times$0.96\arcsec\ ($-68^\circ$)   &      1.6\arcsec$\times$0.91\arcsec\ ($+55^\circ$) \\ \hline
\end{tabular}

$^{a}$Size of Gaussian from fit in $(u,v)$-plane (i.e., deconvolved FWHM size) and position angle of major axis (in parentheses).
\end{table}

\begin{figure}
\resizebox{\hsize}{!}{\includegraphics{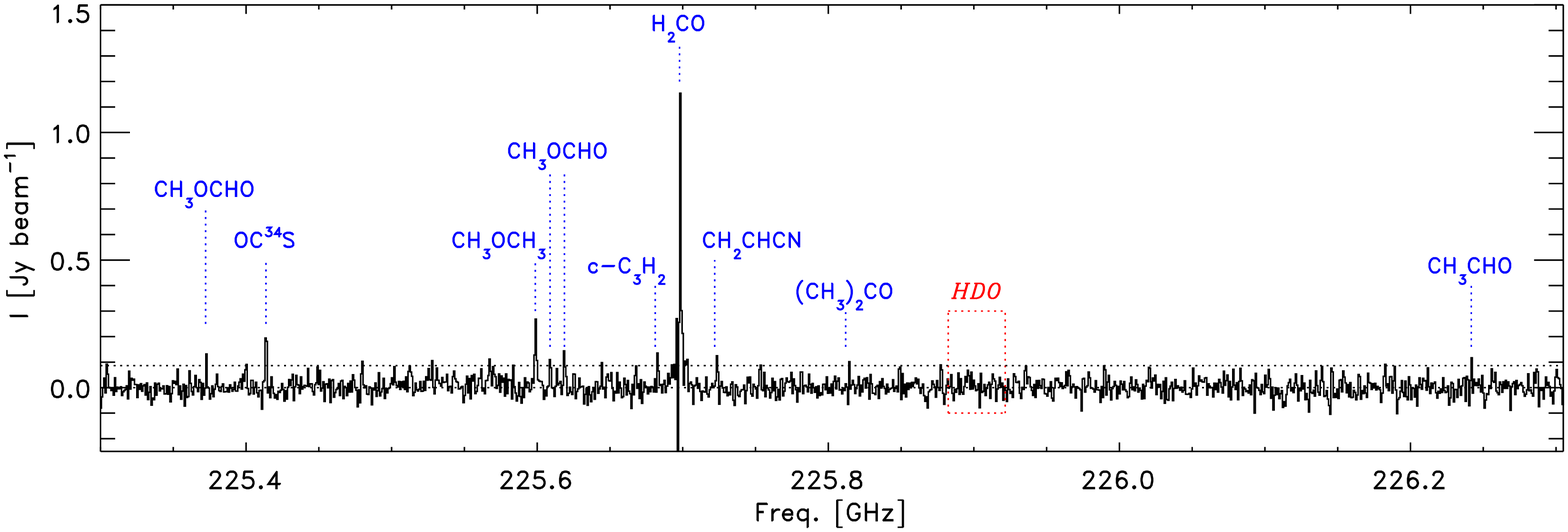}}
\caption{Spectrum extracted in the central $3.0''\times 2.5''$ beam
  toward the continuum position for IRAS4B. The detected lines are
  indicated at the position of their catalog rest frequency corrected
  for the 7.0~\kms\ systemic velocity of IRAS4B. The box shows the
  zoom-in around the HDO $3_{1,2}-2_{2,1}$ transition displayed in
  Fig.~\ref{spectrum_zoom}; the high spectral resolution chunk
  including this line has been rebinned to the same 1.1~\kms\ spectral
  resolution as for the other chunks in the wider bands. The
  horizontal dotted lines indicate the zero- and 3$\sigma$
  levels.}\label{spectrum}
\end{figure}

Fig.~\ref{spectrum} shows the spectrum from 1~GHz of the LSB of IRAS4B
labeling a number of lines with peaks at the 3$\sigma$ level or
above. The H$_2$CO $3_{1,2}-2_{1,1}$ transition at 225.698~GHz is
clearly dominant as expected, with lines of complex organic molecules
as well as OC$^{34}$S and c-C$_3$H$_2$ also detected. Most importantly
the identification of the multiple lines centered at the continuum
position confirms both the pointing of the observations as well as the
calibration of the frequency axes of the spectra.

\begin{figure}
\resizebox{\hsize}{!}{\includegraphics{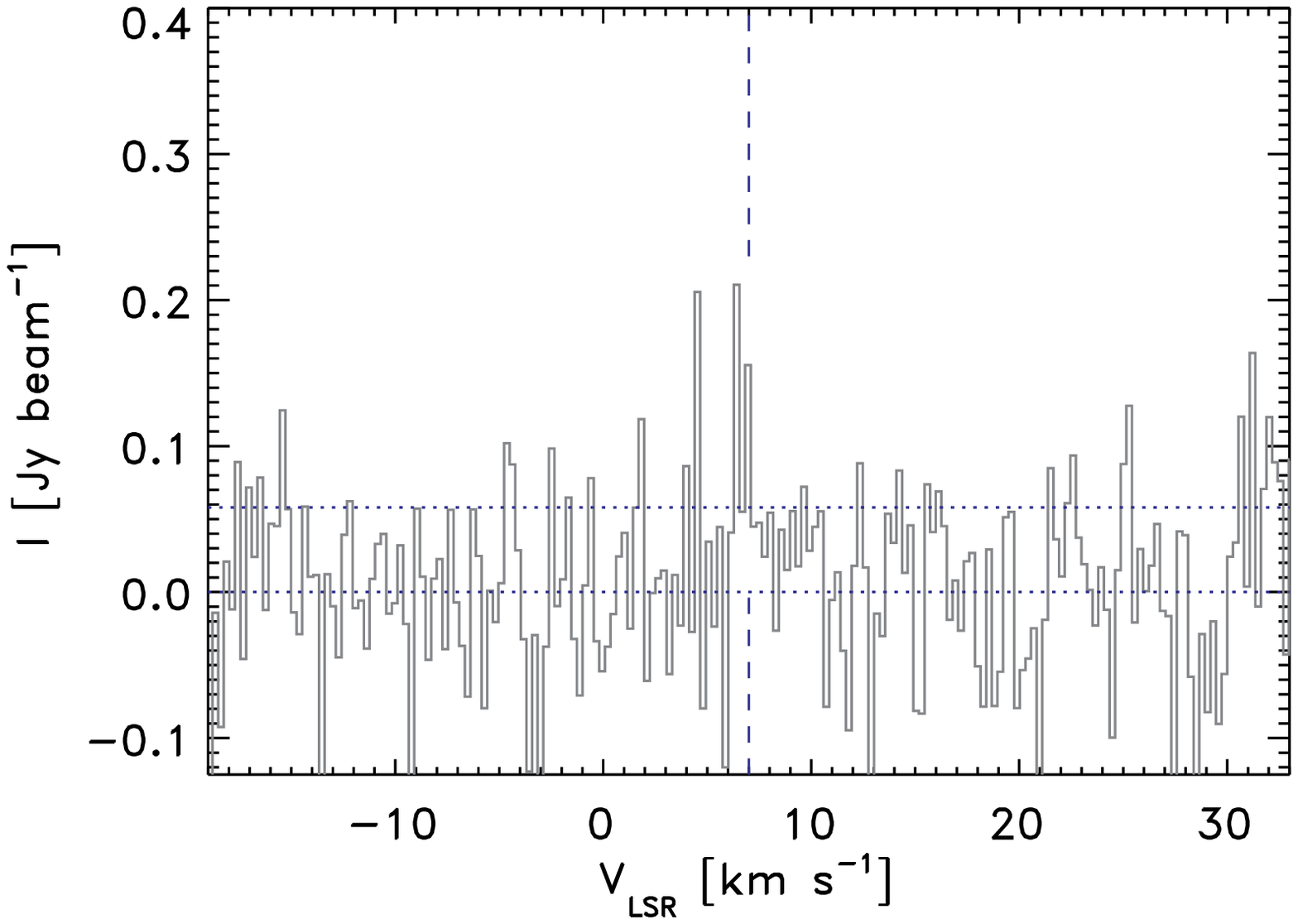}\includegraphics{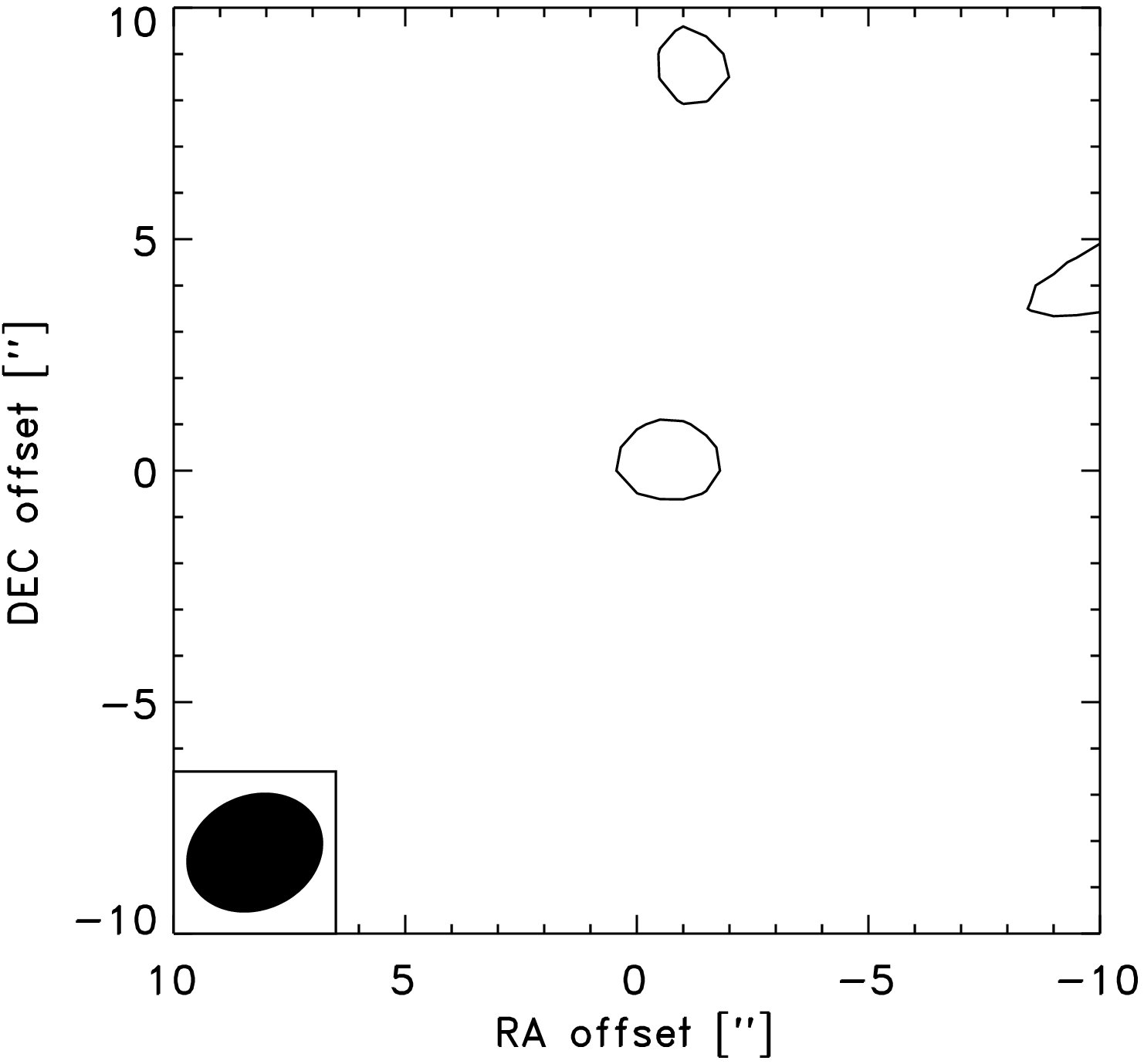}}
\caption{\emph{Left:} Spectrum centered on the expected location of
  the HDO $3_{1,2}-2_{2,1}$ transition. The dotted lines indicate the
  zero- and 1$\sigma$ levels. The dashed line, the expected location
  of the HDO line at the systemic velocity of the H$_2^{18}$O
  $3_{1,3}-2_{2,0}$ transition and other species at
  7.0~\kms. \emph{Right:} Integrated intensity of the data cube over
  $\pm$~1~\kms\ around the expected location of the HDO
  transition. The contour corresponds to the 3$\sigma$ level (no
  emission is seen at 4$\sigma$ levels or
  higher).}\label{spectrum_zoom}
\end{figure}

Fig.~\ref{spectrum_zoom} shows a zoom-in on the spectrum in the high
resolution chunk around the HDO $3_{1,2}-2_{2,1}$ transition. No clear
line is seen at the systemic velocity of 7.0~\kms\ of the H$_2^{18}$O
$3_{1,3}-2_{2,0}$ transition \citep{iras4b_h2o} and other species
surveyed at high angular resolution with the SMA \citep{prosacpaper} -
although a tentative detection may be present at the
$\approx$3$\sigma$ level (integrated intensity). This is confirmed by
the integrated intensity map (Fig.~\ref{spectrum_zoom}, right), which
shows a 3$\sigma$ contour close to the location of the continuum
peak. Statistically speaking it cannot be counted a firm detection,
however, and we therefore treat it as an upper limit in the following.

\section{Discussion: an upper limit on the ratio of the HDO/H$_2$O
  abundances in the inner 50~AU regions of a low-mass protostar}
The $3\sigma$ upper limit on the integrated line intensity (or
approximate strength of the tentative detection) of the HDO transition
can be estimated as $3\sigma_{\rm int} = 3\sqrt{\Delta v\, \delta
  v}\,\sigma_{\rm rms} \approx 0.13$~Jy~\kms, assuming a line width to
zero intensity of 2~\kms\ as found for the H$_2^{18}$O transition. The
column density of HDO corresponding to this intensity can be estimated
under the same assumptions as for the H$_2^{18}$O transition - and can
thus be compared directly, assuming that the emission is optically
thin and uniform over the extent of the H$_2^{18}$O emission (thus
diluted in our SMA beam).

\begin{figure}
  \resizebox{\hsize}{!}{\includegraphics{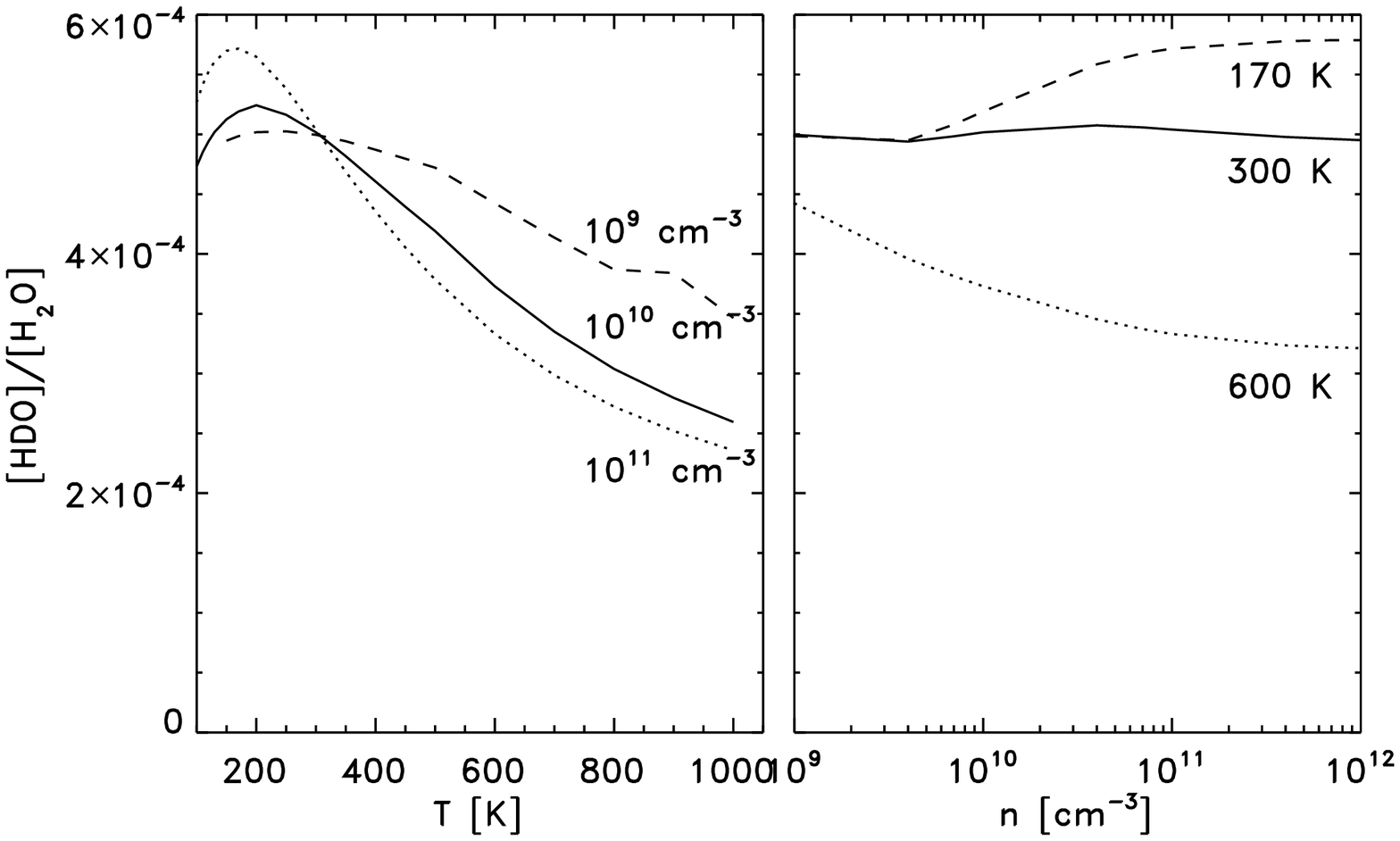}}
\caption{Estimates of the HDO/H$_2$O abundance ratio as function of
  temperature (left) and density (right) from non-LTE Radex
  calculations. Each panel shows three curves: in the left panel for
  constant densities of 10$^{9}$~cm$^{-3}$, 10$^{10}$~cm$^{-3}$ and
  10$^{11}$~cm$^{-3}$ and in the right panel for constant
  temperatures of 170~K, 300~K and 600~K.}\label{radex_calculation}
\end{figure}

Using the non-LTE escape probability code Radex
\citep{vandertak07radex} we calculated column densities corresponding
to the observed line intensities for different values of the kinetic
temperature and H$_2$ density. Fig.~\ref{radex_calculation} shows the
derived HDO/H$_2$O abundance ratio as function of kinetic temperature
and density, respectively. In the range from about 100 to 1000~K and
for densities larger than about $10^{9}$~cm$^{-3}$ (the density
expected in the inner envelope or disk) the derived [HDO]/[H$_2$O]
3$\sigma$ upper limit on the abundance ratio varies from about
2$\times 10^{-4}$ to 6$\times 10^{-4}$ assuming an ortho-para ratio
for H$_2^{18}$O of 3:1 and a $^{16}$O/$^{18}$O ratio of 560. If the
tentative detection would be confirmed by more sensitive observations,
the implied HDO/H$_2$ abundance ratio would suggest a level of
deuteration comparable to what is observed in cometary ices. The Radex
calculations show that the inferred abundance ratio can vary from
about 1$\times 10^{-4}$ to 1$\times 10^{-3}$ if the excitation
temperatures are allowed to differ for the two species and vary
between 100 and 1000~K, covering a wider range of physical
parameters. The beauty of the above result is that it does not depend
on the assumed physical structure of the source - and in particular,
whether the origin of the observed water is due to shocks or ice
desorption in either envelope or disk material.

One caveat is that the H$_2^{18}$O transition could be masing as is
the case for the same transition of main H$_2^{16}$O isotologue at
183~GHz \citep[e.g.,][]{cernicharo94}. In that scenario the
H$_2^{18}$O column density would be overestimated and thus the
[HDO]/[H$_2$O] ratio underestimated. A few observational facts argue
against this being a major issue for the H$_2^{18}$O transition toward
IRAS4B, though: the width of the H$_2^{18}$O transition (2~\kms\ to
zero-intensity) is very close to those of the other species (SO$_2$,
C$_2$H$_5$CN and CH$_3$OCH$_3$) observed in the central beam toward
the continuum peak of IRAS4B. This trend is further strengthened with
detections of the H$_2^{18}$O transition toward two other embedded
protostars (M.~Persson et al., in prep.): the lines of H$_2^{18}$O,
SO$_2$ and the organic species in these three sources show smaller
species-to-species variations than source-to-source variations both in
terms of their line widths and strengths. This suggests that the
origin of the emission is the same in each of the individual sources
and therefore that H$_2^{18}$O emission is dominated by the thermal
component and at most weakly masing. Finally, previously detected
H$_2$O water masers at 22~GHz in IRAS4B are clearly offset from the
systemic velocity \citep[][]{marvel08,desmurs09} and their motions in
the outflow propagation direction are perpendicular to the tentative
velocity gradient observed in the H$_2^{18}$O transition. Thus, the
H$_2^{18}$O line is at least not masing in the gas where the 22~GHz
H$_2$O maser has its origin.

The derived upper limit to the HDO/H$_2$O abundance ratio does not
support the suggestion by \cite{ceccarelli05} and \cite{parise05} that
the HDO/H$_2$O abundance ratio is abnormally high in low-mass
protostars or their disks compared to the abundance ratios found in
comets, Earth's oceans or in high-mass hot core regions
\citep[e.g.,][]{jacq90,gensheimer96,vandertak06}. In agreement with
\cite{stark04}, the results favor a scenario in which the HDO/H$_2$O
ratio is conserved during the star formation process. There is thus
also no need to invoke the formation of water in the gas-phase in a
longer-lived early stage of low-mass protostars where the deuterium
fractionation is high to explain the constraints on the HDO/H$_2$O
abundance ratio. The conclusion that the HDO/H$_2$O abudance ratio is
determined in the cold core phase and is conserved throughout the
formation of low-mass stars and icy solar-system bodies is consistent
with the evolutionary models of \cite{visser09} that show that the
bulk of the ice remains in solid form during infall and disk
formation. It is possible that the physical mechanisms for the origin
of the warm gas in NGC~1333-IRAS4B and IRAS~16293-2422 differ: the
latter source shows a complex structure with multiple organic species
peaking at different locations \citep[e.g.,][]{kuan04,bisschop08} and
likely interaction with the protostellar outflow on small scales
\citep{chandler05}. SMA observations of that source
(J.~K. J{\o}rgensen et al., in prep.) suggests that this complex
interplay could also have an impact on the line emission for the HDO
isotopologue underlining the need for high angular resolution
observations.

In any case, these observations have shown the great potential offered
by high angular resolution submillimeter (interferometric)
observations, even from the ground, for tracing the origin and
evolution of water vapor during the evolution of young stars. An
important step forward would be to spatially resolve both the emission
profiles of HDO and H$_2^{18}$O in protostars in different
evolutionary stages to reveal any temporal or spatial variations of
their abundance ratios. More sensitive observations with the Plateau
de Bure as well as future observations with the Atacama Large
Millimeter Array in Band 5 will make this possible and thereby make an
important contribution toward answering the questions about the origin
of water in planetary systems.

\acknowledgments 

We are grateful to Tim van Kempen for helpful comments about the
manuscript. Research at Centre for Star and Planet Formation is funded
by the Danish National Research Foundation and the University of
Copenhagen's programme of excellence. Research in astrochemistry in
Leiden is supported by a Spinoza Grant from the Netherlands
Organization for Scientific Research (NWO) and a NOVA grant. This
paper is based on data from the Submillimeter Array: the Submillimeter
Array is a joint project between the Smithsonian Astrophysical
Observatory and the Academia Sinica Institute of Astronomy and
Astrophysics and is funded by the Smithsonian Institution and the
Academia Sinica.

\end{document}